# Multi-fibers connectors systems for FOCCoS-PFS-Subaru

Antonio Cesar de Oliveira[1], Ligia Souza de Oliveira[2], Lucas Souza Marrara[2], Leandro Henrique dos Santos[1], Marcio Vital de Arruda[1], Jesulino Bispo dos Santos[1], Décio Ferreira[1], Josimar Aparecido Rosa[1], Rodrigo de Paiva Vilaça[1], Laerte Sodré Junior[3], Claudia Mendes de Oliveira[3], James E. Gunn[4].

1- MCT/LNA –Laboratório Nacional de Astrofísica, Itajubá - MG – Brazil
2- OIO-Oliveira Instrumentação Óptica LTDA – SP - Brazil
3- IAG/USP – Instituto de Astronomia, Geofísica e Ciências Atmosféricas/ Universidade de São Paulo - SP – Brazil
4- Department of Astrophysical Sciences - Princeton University - USA

## ABSTRACT

The Fiber Optical Cable and Connector System (FOCCoS), provides optical connection between 2400 positioners and a set of spectrographs through optical fibers cables as part of PFS instrument for Subaru telescope. The optical fiber cable will be segmented in 3 parts along the route, cable A, cable B and cable C, connected by a set of multi-fiber connectors. The company USCONEC produces the multi-fiber connector under study. The USCONEC 32F model can connect 32 optical fibers in a 4 x 8 matrix arrangement. The ferrules are made of a durable composite, Polyphenylene Sulfide (PPS) based thermoplastic. The connections are held in place by a push-on/pull-off latch, and the connector can also be distinguished by a pair of metal guide pins that protrude from the front of the connector. Two fibers per connector will be used for monitoring the connection procedure. It was found to be easy to polish and it is small enough to be mounted in groups.

Highly multiplexed instruments like PFS require a fiber connector system that can deliver excellent optical performance and reliability. PFS requires two different types of structures to organize the connectors. The Tower Connector system, with 80 multi-fiber connectors, will be a group of connectors for connecting cable B (Telescope Structure) with cable C (Positioners Plate). The Gang Connector system is a group of 8 gang connectors, each one with 12 multi-fibers connectors, for connecting cable B (Telescope Structure) with cable A (Spectrograph). The bench tests with these connector systems and the chosen fibers should measure the throughput of light and the stability after many connections and disconnections. In this paper we describe tests and procedures to evaluate the throughput and FRD increment. The lifetime of the ferrules is also in evaluation.

## 1. INTRODUCTION

The FOCCoS sub-system of PFS is comprised of 3 optical fiber cables, which are responsible for guiding light from 2394 positioners to 4 spectrographs.[01,02] The three cables are joined by two connector assemblies: 1) the tower connector, which is located on the telescope's spider vanes, and 2) the Gang connector, which is located close to the spectrographs.. The Tower connector is necessary to enable removal of the PFI from the Subaru telescope structure, allowing exchange of instruments. The Gang connector is necessary for logistics reasons, such as possible improvements of the instrument without removal of all fibers between the spectrographs and the top end. Between the two-connector assemblies, cable B will be a permanent cable installed as part of the structure of the telescope. However, the ability to connect and disconnect cable B from the spectrographs offers facilities to explore improvements modularly.

## 1.1 Ferrule multi-fibers connector

The US Conec brand MTP connector is based on mechanical transfer ferrule technology. The MTP® (mechanical transfer push-on), shown in Figure 1, is US Conec's trademarked enhanced-performance version of the MPO (multi-fiber push-on connector). The MTP® ferrule is a monolithic, high precision, molded component; custom ferrules can be made to order for a given fiber count and fiber outer diameter. The ferrules are made of Polyphenylene Sulfide (PPS) based thermoplastic. The connections are held in place by a push-on/pull-off latch, and the connector can also be distinguished by a pair of metal guide pins that protrude from the front of the connector. Ferrule 32F, shown in Fig. 2, will be the cell of the connector benches for this project. It is easy to mount, easy to polish and small enough to be mounted in groups. Although it has been used in the Apogee spectrograph (SDSS) with success, the lifetime and throughput of this connector are being investigated in the context of PFS development.

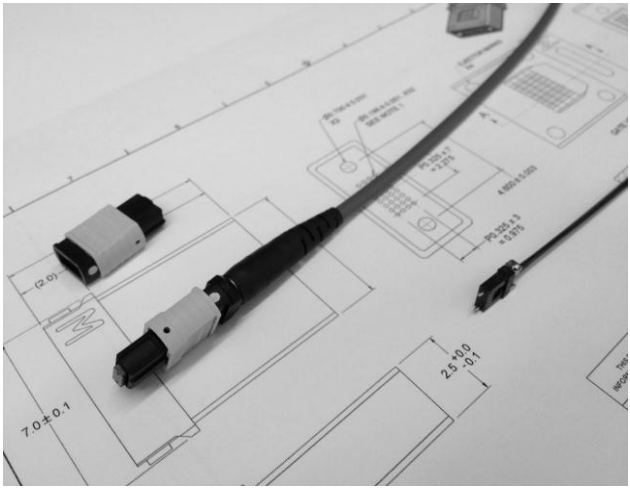

**Figure 1** – US Conec MTP multi-fibers connector produced by US Conec Company.

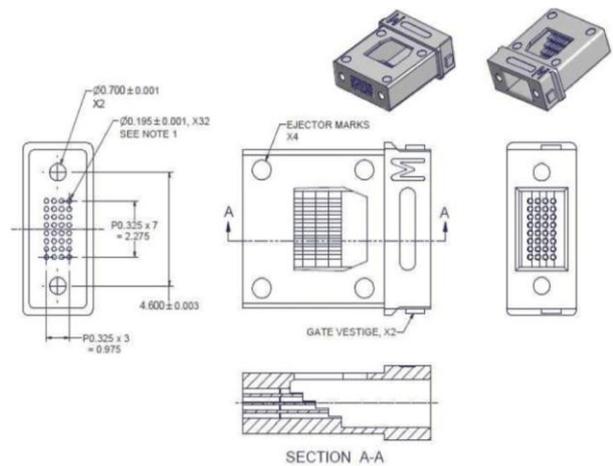

**Figure 3 -** Two possible fibers arrangement inside the US Conec Ferrules, planned to receive 57 science fibers. Two monitoring fibers

US Conec ferrule 32F has potentially 32 channels to insert optical fibers with 190um, distributed in a 4x8 matrix, as shown in Fig. 3. Possible fiber arrangements inside the Ferrules include 29 or 28 fibers, enabling the necessary fiber distribution. Monitor fibers can be inserted in the diagonal extremities of each ferrule. The analysis of the light flux from these fibers gives direct information about the connection per ferrules pair. In fact, this is the best position to evaluate the connection taking in account the parallelism between the opposing ferrule surfaces. If both monitoring fibers have good connection, all fibers of the ferrule also should have good connection.

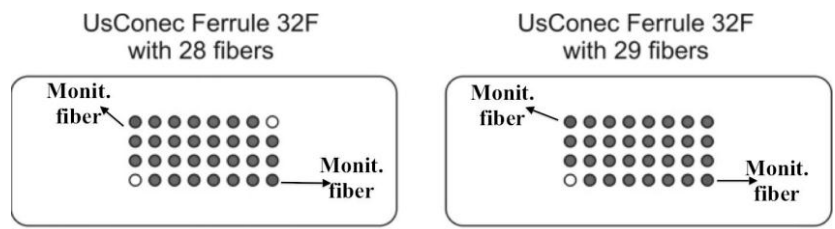

**Figure 2 -** Schematic of ferrule 32F. Information provided by the US Conec Company (www.usconec.com).

**1.2 Fiber optics connection**

The US Connec MTP multi-fiber connector has a spring mechanism that can be used to define a range of force against the contact. This force is important to ensure a contact pressure between the two sides of the connector and to maintain a stable connection. This contact pressure between the surfaces of the fibers shapes the interface of the optical fibers during the connection, enabling a higher coupling efficiency. Thus, this type of connector defines a dynamic range of opposing force to improve the contact interface, therefore ensuring better performance of the connection.

## 2.0 OPTICAL BENCH CONNECTORS

FOCCoS sub-system of PFS has two optical fiber bench connectors. Both benches use the US Conec MTP as structural cells in a design that exploits the contact force in each individual cell. The design uses the precision from the US Conec ferrule 32 F and the high performance of the fibers in the project. Considering the limited space at the top end of the telescope, a completely new bench for the USCONEC ferrule – the Tower Connector – was developed to connect cables C and B. However, we do not have the same limitation of space to connect the cables B and A at the IR-side fourth floor, so a well-known connector bench, Gang Connector, will be used.

**2.1 Gang connector**

Gang connector was developed for the APOGEE spectrograph [03] and designed by R. French Leger (Washington University). This bench, illustrated by Figs. 4 and 5, can accommodate 10 or 12 US Conec multi-fibers connectors. In this way, the optical fibers from each sub-cable of Cable B will be distributed in one Gang connector (male side). The female side is the beginning of the Cable A, which leads to the spectrographs. The Gang connector will be installed at the IR - side fourth floor. The number of cells requires around 20 kgf to balance the sum of the individual strength of each spring from each US Conec connector. The process to ensure that force includes a mechanism to obtain displacement of the core with subsequent generation of opposing forces between pairs of connectors, using a simple rotation of the connector outer shield.

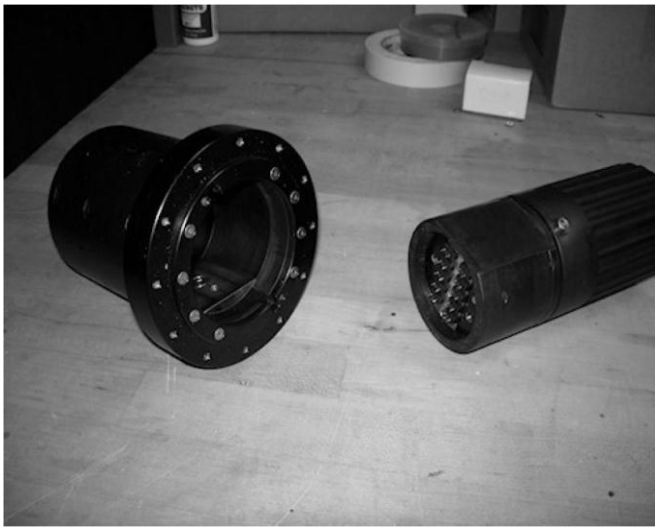
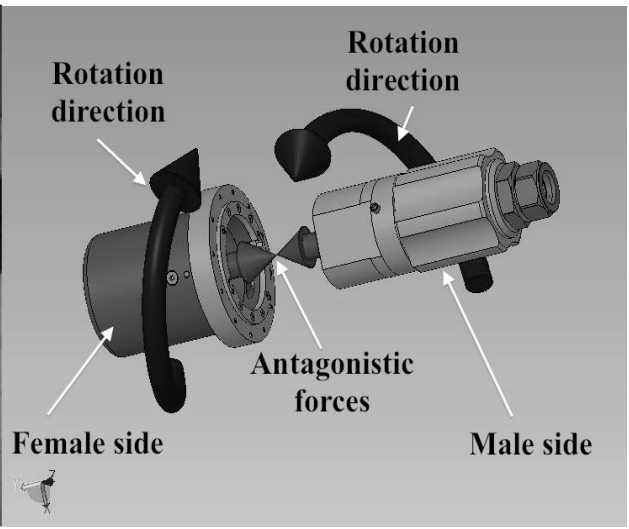

**Figure 4:** Gang connector system, to be used at spectrograph side of Cable B. 4 Gang connectors with 10 cells and 4 Gang connectors with 11 cells compose the bench.

**Figure 5:** Drawing of Gang connector showing the Male side, the female side and the opposing forces to ensure the pressure of contact between the ferrules.

## 2.2 Tower connector

The Tower connector is the connector bench responsible for the connection of the optical fibers from cable C to cable B. It was designed to fit the available space at the spider of the telescope, as illustrated by Figs. 6 and 7. The male side of the connection is part of cable C; the female connection is part of cable B. The core of the Tower Connector is the already described US Conec multi-fibers connector. Each cell is populated with 28 or 29 optical fibers. Therefore, each two cells in the Tower Connector will connect the 57 science fibers from a single COBRA module assembly distributed, Fig. 8. The 'connector unit' of the Tower Connector contains 22 cells, as can be seen in Figure 22. This number of cells requires around 38 kgf to balance the sum of the individual strength of each spring from each US Conec connector. The procedure to obtain this force during the connection uses a lever system, Fig. 9.

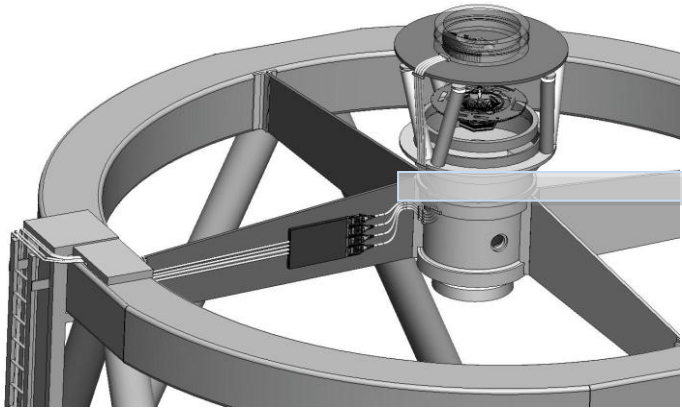

**Figure 6:** Tower connector bench installed at the spider on the top end of Subaru telescope.

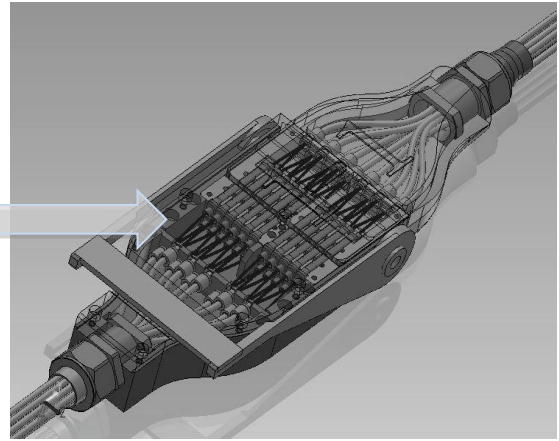

**Figure 7:** Tower connector bench is composed by 4 Tower connectors unity like is shown in this drawing.

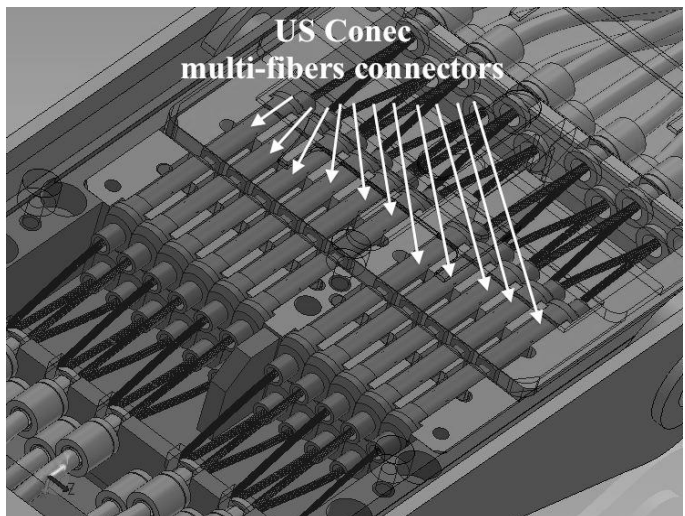

**Figure 8:** Internal details of the Tower connector unity with 22 US Conec multi-fibers connectors.

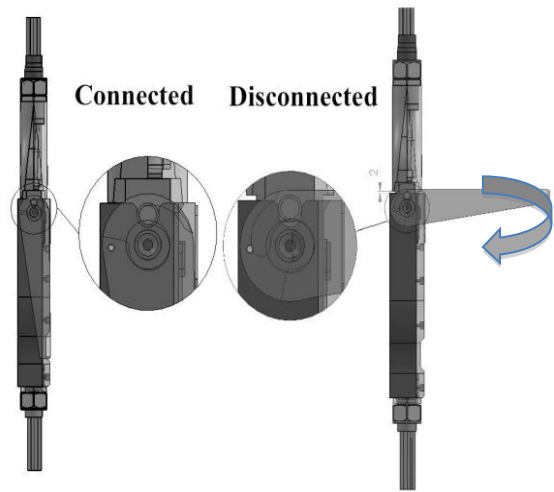

**Figure 9:** Process to convert rotation in displacement with opposing forces between the sides of the connectors

# 3. TESTING MULTI-FIBERS CONNECTORS

Although there is some suitable professional equipment to measure coupling efficiency of fiber optic connectors, we are interested in measuring efficiency of connection considering the propagating energy profile to the optical fiber and from the optical fiber. This paper presents a new approach to evaluate the rate of transfer of light between either side of the connector, which takes into account the focal reasons involved in the assembly under test. This approach defines the perspective to obtaining a signature of a connection, which can eventually help to lifetime testing of the connector.

## 3.1 Transfer rate test

To measure the transfer rate of the multi-fibers connectors we have used the experimental apparatus illustrated in Fig. 10. It illuminates the test fiber with an f/2.8 beam and computes the ratio between the energy encircled by selected f/# in the fiber output beam to the total energy of the input beam. The pinhole P (0.8 mm) is illuminated by a beam coming from a stabilized halogen lamp.

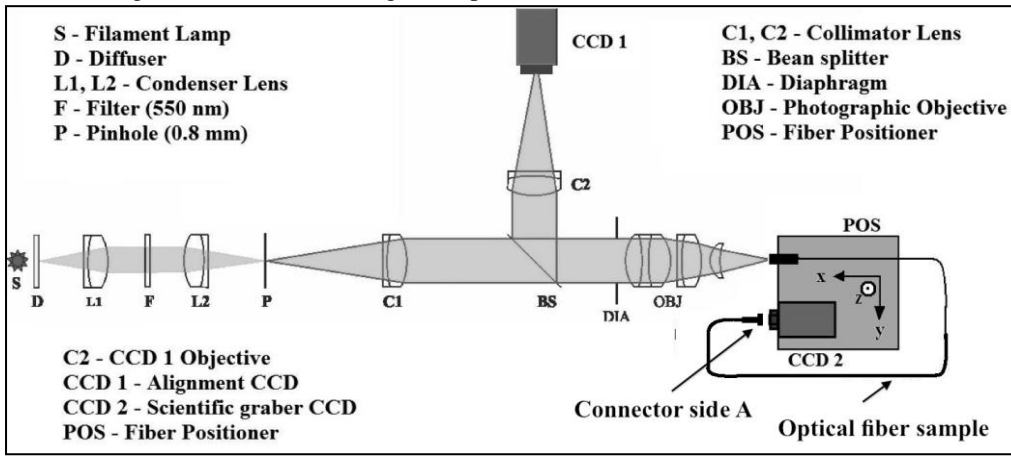

**Figure 10** – First part of the procedure to measure the transfer rate between the connections.

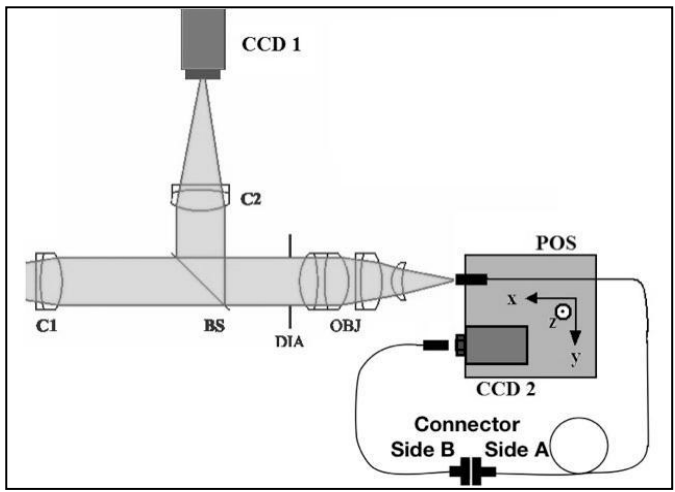

**Figure 11:** Second part of the procedure to measure the transfer rate between the connections.

The uniformity of the beam and its predominant wavelength are obtained using a diffuser D and an interference filter F centered at $\lambda = 550 \pm 10$nm. The main optical system, composed of an achromatic doublet C1 (f = 400 mm) and a photographic lens OBJ (f = 50 mm), produces a 8X de-magnified pinhole image on the measurement plane (MP). An optical subsystem, formed by the lens C2 and a commercial grade camera CCD1, monitors the position and alignment of the optical fiber on MP. This subsystem is connected to the optical axis of the main system through the beam splitter BS. A diaphragm DIA placed at the front focal plane of OBJ controls the beam focal ratio. The detector is a scientific grade camera CCD2 placed at the back focal point of an achromatic doublet L4 (f = 12.7 mm). The front focal point of L4 lens is called the focal point of the detector. Initially the system is configured to take images from the side A of the connector as we can see in the Fig.10. The fiber sample entrance is positioned on the image of the pinhole formed in the measurement plane MP and aligned with the optical axis of the main system by means of a three-axis platform. A reference image is taken in that position, and also a background image. Then the system is configured to take images from the side B of the connector, Fig. 11. The other side of the fiber is positioned and aligned with the detector focal plane. New image is taken as well as a background image. Comparing the reference image with the image data we can determine the transfer rate of energy transmitted by the fiber as well as its spatial distribution of encircled energy.

**3.2 Software and Calculations**

To evaluate the transfer rate of the connector with fiber submitted a particular Input Focal Ratio, the software takes the concentric rings centered on the fiber image, from exit side A and from exit side B. This is used to define the efficiency over a range of f-numbers at the exit of the fiber, side A and B, where each f-number value contains the summation of partial energy emergent from each fiber side.

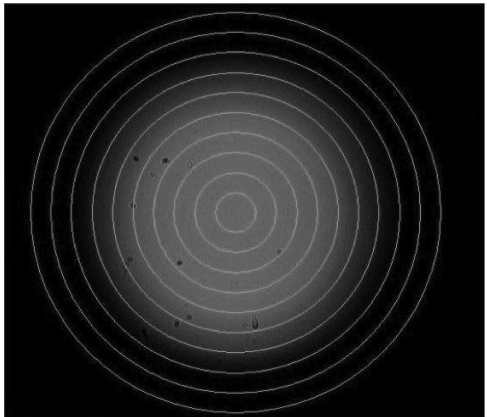

Encircled energy by several rings in the image of the spot light from the fiber exit, side A

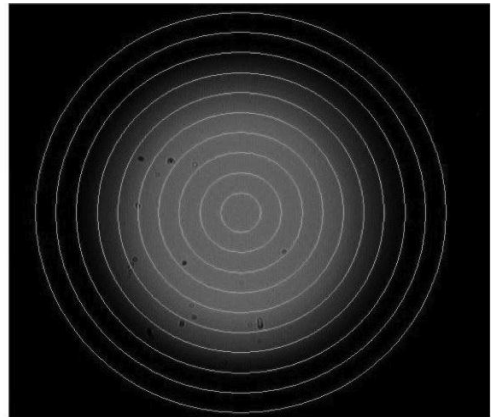

Encircled energy by several rings in the image of the spot light from the fiber exit, side B

$$E_{O\,B\,Encircled\,i} \div E_{O\,A\,Encircled\,i} = Transfer\ ratio$$

**Figure 12:** Description of the methodology adopted to measure the transfer rate of the connector. The mathematical analysis shows that it is not necessary take any image from the light source.

The limiting focal ratio that can propagate in the tested fiber is approximately F/2.2 taking in account the numerical aperture, (NA), of this fiber to be $0.22 \pm 0.02$. Therefore, we have defined F/2.2 to be the outer limit of the external annulus within which all of the light from the test fiber will be collected. The corresponding diameters of the rings are converted to output focal ratios, multiplying them by the appropriate constant given by the distance between the fiber output end and the detector. To obtain the transfer rate curve, we just need divide B/A. In this case, each ratio value is calculated by the number of counts within each annulus (i) from the image of the fiber exit side A, divided by number of counts within the corresponding annulus (i) from the image of the fiber exit side B. This process gives us a direct ratio between the encircled energy, from the connector side B with that comes from the connector side A, Fig.12.

**3.3 Samples in test**

We have worked with a set of 4 samples, each one with multi-fiber connectors in both extremities populated with 32 optical fibers 1-meter length. One pair of samples was made with Fujikura fibers, (130170191) and the other pair was made with Polymicro fibers, (FBP120170190). FOCCoS project use fibers from these two manufactures, so tests with the connectors need to be made with both types of fibers. In fact, the efficiency of the connection depends strongly on the combination between connector and optical fiber. In this context we have done 3 sets of measurements; Polymicro connected with Polymicro, Fujikura connected with Fujikura and Polymicro connected with Fujikura. For each one of those combinations it is important to measure the 32 fibers from the connected cables to evaluate the signature of the connection. Any test needs to be made with some pressure of contact so we have constructed a special structure to ensure the opposing contact forces between both sides of the connector, Fig. 13.

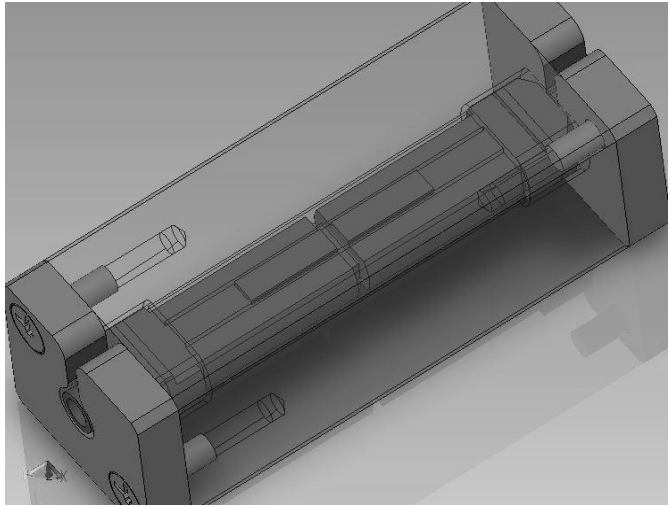

**Figure 13:** Drawing of the device that keeps both sides of the multi-fibers connector coupled with some contact pressure. The design also keeps both sides mechanically aligned.

## 4. RESULTS

We present in this section some results obtained from the measurements of transfer rate described in the sub-section 3.1 and 3.2 with the samples described in sub-section 3.3. These tests produce an enormous amount of data; so it is necessary make some simplifications to get some understanding.

Fig. 14 shows a plot of 32 curves of transference rate obtained from the connection of Polymicro against Polymicro fibers. This plot shows us the behavior of the energy profile from each fiber of the sample after the connection as a function of the Numerical Aperture, (Focal Ratio). Evaluation of a large numbers of fibers per multi-fibers connector is complicated by small variation that can be caused by experimental errors or even minor damage to the surface caused by mechanical wear during connection and disconnection operations. Although we are interested in evaluating fiber by fiber, it is more interesting to evaluate a complete connector. Thus, we can obtain a simple average by adding all the curves and dividing by the number of fibers. This simple average produces a single curve showing the performance of the connector as a whole. Fig. 15 shows a plot of 2 curves, each of which represents a simple average of 32 curves previously obtained.

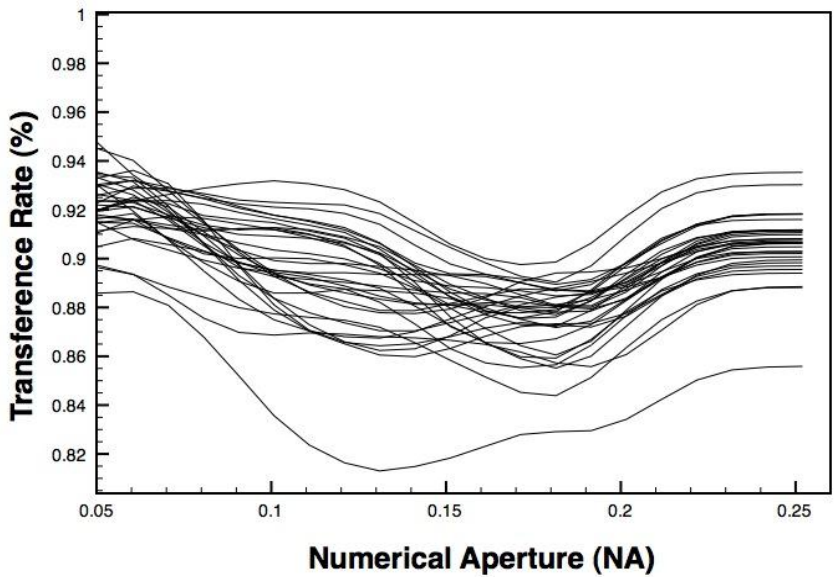

**Figure 14:** Connection between 32 Polymicro fibers

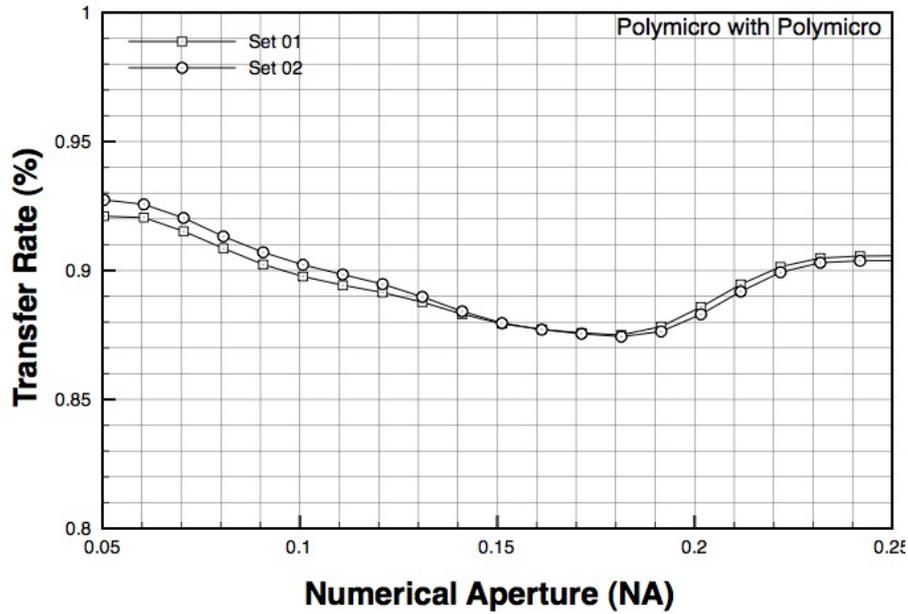

**Figure 15:** The figure shows curves obtained from the average between 32 connections, Polymicro against Polymicro fibers. Two sets of measurements were obtained with an interval time of 1 day including the operation to disconnect and connect again. The variations between both curves are less than 1% and the experimental error was calculated to be around 4%. The entrance focal ratio was fixed in F/2.8 - (NA=0.178)

The analysis of Fig. 15 clearly shows a curve that defines a signature connection. The same can be observed in the Fig. 16, which shows the signature of connection between Fujikura fibers. The Fig. 17 shows the signature of connections between Polymicro and Fujikura fibers. In this case we had care to define the Polymicro cable side the entrance of light. This was necessary taking into account that the Polymicro fiber used in this experiment has a core diameter smaller than the Fujikura core diameter, by 10 microns.

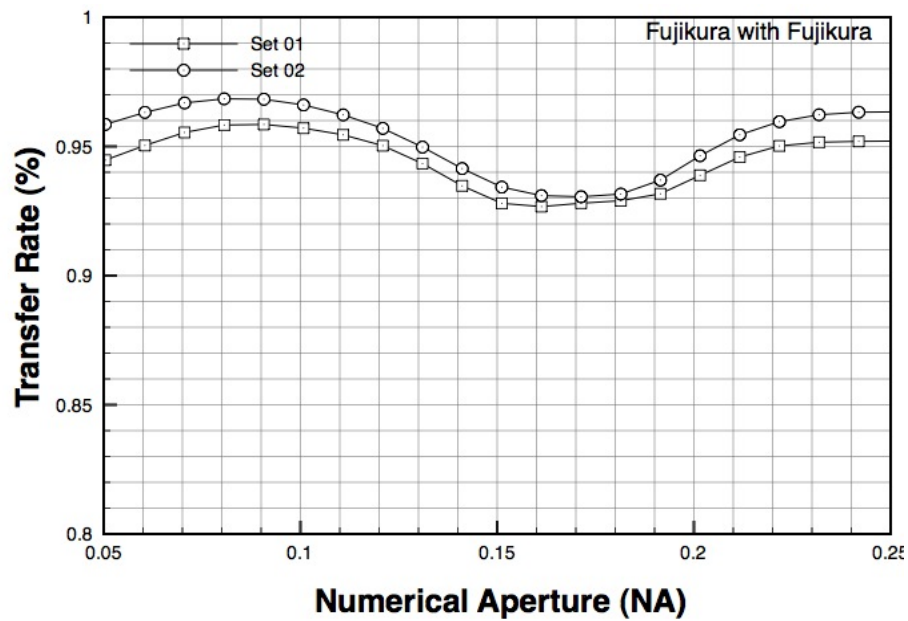

**Figure 16:** The figure shows curves obtained from the average between 32 connections, Fujikura against Fujikura fibers. Two sets of measurements were obtained with an interval time of 1 day including the operation to disconnect and connect again. The variations between both curves are less than 1% and the experimental error was calculated to be around 4%. The entrance focal ratio was fixed in F/2.8 - (NA=0.178)

This is important to optimize the light transference in the interface. In fact the performance looks better when the light goes from smallest to largest core diameter.

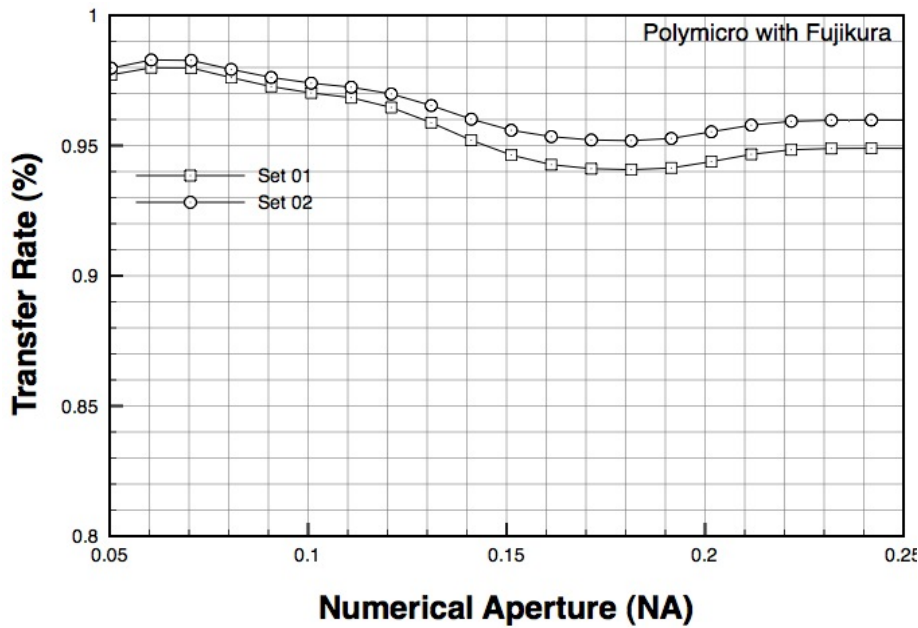

**Figure 17:** The figure shows curves obtained from the average between 32 connections, Polymicro against Fujikura fibers. Two sets of measurements were obtained with an interval time of 1 day including the operation to disconnect and connect again. The variations between both curves are around 1% and the experimental error was calculated to be around 4%. The entrance focal ratio was fixed in F/2.8 - (NA=0.178)

## 5. SUMMARY AND CONCLUSIONS

We described in this paper the connector system that will be used in FOCCoS/ PFS project to ensure the connection of 3 sets of cables within 2394 optical fibers. This connector system is comprised of two bench multi-fibers connectors, called the Gang connector and the Tower connector. The Gang connector has been used with success in the APOGEE spectrograph project for Apache Point Observatory. The Tower connector required a new design because of space restrictions on the telescope's spider vane. The basic core of both systems is the US Conec multi-fiber connector, which is our object under study in this work. We present here a new concept to measure and evaluate the efficiency of connection with initial results that can predict the existence of "*connection signature*" for each set of connector plus optical fiber. This signature of connection can be extremely useful for evaluating the lifetime of connectors like this, taking into account the possible mechanical damage caused by the successive processes of connection and disconnection.

## 6. ACKNOWLEDGMENTS

We gratefully acknowledge support from: Fundação de Amparo a Pesquisa do Estado de São Paulo (FAPESP), Brasil. Laboratório Nacional de Astrofísica, (LNA) e Ministério da Ciência Tecnologia e Inovação, (MCTI), Brasil. We would like also to gratefully acknowledge French Leger (Washington University) for help us with the theme.

## 7. REFERENCES

[01] Sugai, H., et al., "Prime focus spectrograph: Subaru's future" Proc. SPIE 8446, Ground-based and Airborne Instrumentation for Astronomy IV, 84460Y (2012)
[02] Oliveira, A. C., et al., "FOCCoS for Subaru PFS" Proc. SPIE 8446, Ground-based and Airborne Instrumentation for Astronomy IV, 84464R (2012)
[03] Wilson, J. C., et al., "The Apache Point Observatory Galactic Evolution Experiment (APOGEE) high-resolution near-infrared multi-object fiber spectrograph" Proc. SPIE 7735, Ground-based and Airborne Instrumentation for Astronomy III, 77351C (2010)